\documentclass[conference]{IEEEtran}

\usepackage{graphicx}
\usepackage{latexsym}

\usepackage{listings}
\usepackage{tabularx}

\def\Underline{\setbox0\hbox\bgroup\let\\\endUnderline}
\def\endUnderline{\vphantom{y}\egroup\smash{\underline{\box0}}\\}
\def\|{\verb|}

\begin{document}

\title{Exploring the ChatGPT Approach for Bidirectional Traceability Problem between Design Models and Code}

\author{\IEEEauthorblockN{Hideyuki Kanuka, Genta Koreki, Ryo Soga, Kazu Nishikawa}
\IEEEauthorblockA{Research \& Development Group, Hitachi, Ltd.}}

\maketitle

\begin{abstract}
This study explores the capabilities of Large Language Models, particularly OpenAI's ChatGPT, in addressing the challenges associated with software modeling, explicitly focusing on the bidirectional traceability problem between design models and code. The objective of this study is to demonstrate the proficiency of ChatGPT in understanding and integrating specific requirements into design models and code. We also explore its potential to offer solutions to the bidirectional traceability problem through a case study. The findings indicate that ChatGPT is capable of generating design models and code from natural language requirements, thereby bridging the gap between these requirements and software modeling. Despite its limitations in suggesting a specific method to resolve the problem using ChatGPT itself, it exhibited the capacity to provide corrections to be consistent between design models and code. As a result, the study concludes that achieving bidirectional traceability between design models and code is feasible using ChatGPT.
\end{abstract}

\section{Introduction}
Large Language Models (LLM) have significantly impacted society in recent years. One prime example is OpenAI's ChatGPT \cite{b1}, which uses natural language processing to generate human-like conversations. Such products have a wide range of applications in various fields.

In recent years, software engineering has seen a surge of interest in applying LLM \cite{llm}. Various LLMs have been used in innovative development methods such as code generation and execution trace analysis to improve coding assistance \cite{barke} and fault diagnosis effectiveness \cite{sakura}. Despite this, research in software modeling, a sub-field of software engineering, remains limited. Reports on research such as the one conducted by Camara et al. on the effectiveness of ChatGPT \cite{camara} and the Unified Modeling Language (UML) are still scarce.

In this study, we aim to explore the potential of ChatGPT, which is based on the LLM GPT-4 \cite{gpt4}, to address the challenges that arise in software modeling. Our focus is on the problem of bidirectional traceability in software engineering, which refers to the ability to track requirements, design, code, and test artifacts in both directions \cite{trace}. Specifically, we concentrate on bidirectional traceability between design models and code in software modeling \cite{eramo}. This ensures consistency and synchronization between design and code as any changes made to one are reflected in all other design models, and vice versa.

The following case study aims to use ChatGPT to generate design models and code that reflect specific requirements. This allows us to demonstrate ChatGPT's ability to comprehend and incorporate these requirements into design models and code. Simultaneously, we present a bidirectional traceability problem to ChatGPT and assess its ability to provide solutions. Furthermore, we intentionally change some parts of a design model and a code to determine how ChatGPT can synchronize them. Through this approach, we assess the effectiveness of using ChatGPT in bidirectional traceability and report our findings in this paper.

\section{Case Study}

For our case study on the bidirectional traceability problem, we have selected the ``Library Problem 2.0" \cite{ubayashi} created by Ubayashi and Kamei. This problem was originally suggested to be used as a common problem in the software engineering community. However, as the original text was in Japanese, we have translated it into English and added some text to make it suitable for use in ChatGPT prompts.

The ``gpt-4-0613" model of GPT-4 is used in the environment for using ChatGPT, and the temperature is set to 0 to minimize the variability in the responses.

\begin{figure}[t]
    \begin{tabularx}{\linewidth}{|X|}
    \hline
    \small
    \textbf{\#Problem:}
    \\
    \small
    Consider a book lending system in a library. The library is open-shelf, and users select books themselves, bring the books they want to borrow to the counter, attach a user card, and apply for lending. The counter staff registers the lending information and returns the book and user card to the user. When returning, the user requests a return at the counter with the book and user card. The counter staff performs the return process and returns the user card. Every day, the counter staff checks the lending status and urges users who have been delayed for more than two weeks by phone.
    \\
    \\
    \small
    \textbf{\#Instruction:}
    \\
    \small
\textbf{For the above \#problem,} create the design model \textbf{in PlantUML format} and the code \textbf{in Python language in detail} and present a method \textbf{by using ChatGPT} to ensure bidirectional traceability between them. The traceability refers to the situation where when the model is changed, the corresponding code is changed in sync, and vice versa.
	\\
    \hline
    \end{tabularx}
\caption{Prompt on a bidirectional traceability problem}
\label{fig:prompt1}
\end{figure}

\subsection{Testing ChatGPT's Ability to Solve the Bidirectional Traceability Problem}

We designed a prompt shown in Figure \ref{fig:prompt1}, which is based on the ``Library Problem 2.0" \cite{ubayashi} to test ChatGPT's ability to solve the bidirectional traceability problem between design models and code. We added bold sections to the original text, including ``\#Problem" and ``\#Instruction" to provide clear instructions to ChatGPT. We specified the format of a ``design model" as PlantUML \cite{plantuml} and a ``code" as Python to obtain specific responses. We also added the phrase ``by using ChatGPT" to evaluate if it could provide a solution to the bidirectional traceability problem.

A class diagram was generated using PlantUML as a design model, which accurately reflected the content of the problem as a requirement. Python language code was also generated. The details of each are shown in Figures \ref{fig:class1} and \ref{fig:python1}. Additionally, Figure \ref{fig:class_diagram} shows an image of the class diagram generated from the PlantUML in Figure \ref{fig:class1}.

Figure \ref{fig:answer1} shows the response to the ``\#Instruction" on bidirectional traceability using ChatGPT. While the response appears correct at first glance, the statement ``We can train ChatGPT..." lacks specificity and does not offer a precise approach. Various methods exist to train ChatGPT, but they can prove challenging for users. For instance, the LLM GPT model can be trained through fine-tuning, but the current GPT-4 model typically cannot undergo fine-tuning. Moreover, even if fine-tuning were possible, it necessitates considerable training data. Another method is few-shot learning, which employs a prompt engineering technique \cite{fewshot}. However, it remains unclear from the response in Figure \ref{fig:answer1} what precisely constitutes a few-shot and how to implement it.

\begin{figure}[t]
\begin{tabularx}{\linewidth}{|X|}
\hline
\small
\#Design Model in PlantUML:
\\
\small
(the design model shown in Figure \ref{fig:class1})
\\
\\
\small
\#Python Code:
\\
\small
(the python code shown in Figure \ref{fig:python1})
\\
\\
\small
\#Bidirectional Traceability:
\\
\small
To ensure bidirectional traceability, we can use ChatGPT to create a mapping between design models and the code. For example, we can train ChatGPT to understand that the `Library` class in the design model corresponds to the `Library` class in the Python code and vice versa. Similarly, the `borrowBook` method in the `Library` class in the design model corresponds to the `borrowBook` method in the `Library` class in the Python code, and so on. This way, if the design model is changed, ChatGPT can automatically generate the corresponding changes in the Python code and vice versa.
\\
\hline
\end{tabularx}
\caption{ChatGPT response to the prompt shown in Figure \ref{fig:prompt1}}
\label{fig:answer1}
\end{figure}

\begin{figure}[h]
\lstset{
    frame=single,
    tabsize=1,
    basicstyle=\ttfamily\footnotesize,
    frame=single,
    breaklines=true
}
\begin{lstlisting}[language=Python]
@startuml
class Library {
  +borrowBook(user: User, book: Book):void
  +returnBook(user: User, book: Book):void
  +checkLendingStatus(): void
}

class User {
  -name: String
  -userCard: UserCard
  +User(name: String)
  +getName(): String
  +getUserCard(): UserCard
}

class UserCard {
  -userID: String
  +UserCard(userID: String)
  +getUserID(): String
}

class Book {
  -title: String
  -borrowed: boolean
  +Book(title: String)
  +getTitle(): String
  +isBorrowed(): boolean
  +setBorrowed(borrowed: boolean): void
}

Library "1" -- "many" User : has
Library "1" -- "many" Book : has
User "1" -- "1" UserCard : has
@enduml
\end{lstlisting}
\caption{A design model in PlantUML format obtained from the prompt in Figure \ref{fig:prompt1}}
\label{fig:class1}
\end{figure}

\begin{figure}
    \centering
    \includegraphics[width=8.5cm]{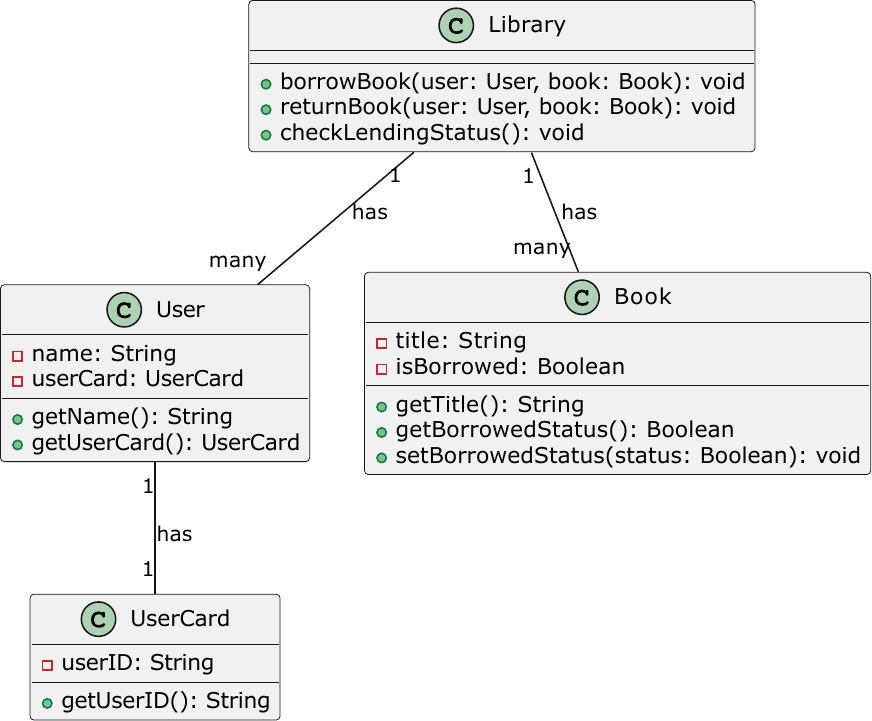}
    \caption{An image of the class diagram generated from Figure \ref{fig:class1}}
    \label{fig:class_diagram}
\end{figure}

\begin{figure}[t]
\lstset{
    frame=single,
    tabsize=1,
    basicstyle=\ttfamily\footnotesize,
    frame=single,
    breaklines=true
}
\begin{lstlisting}[language=python]
class Library:
    def __init__(self):
        self.users = []
        self.books = []
    def borrowBook(self, user, book):
        if book in self.books 
        and not book.isBorrowed():
            book.setBorrowed(True)
            print(f"{user.getName()} has borrowed {book.getTitle()}")
    def returnBook(self, user, book):
        if book in self.books 
        and book.isBorrowed():
            book.setBorrowed(False)
            print(f"{user.getName()} has returned {book.getTitle()}")
    def checkLendingStatus(self):
        for book in self.books:
            if book.isBorrowed():
                print(f"{book.getTitle()} is borrowed")

class User:
    def __init__(self, name, userCard):
        self.name = name
        self.userCard = userCard
    def getName(self):
        return self.name
    def getUserCard(self):
        return self.userCard

class UserCard:
    def __init__(self, userID):
        self.userID = userID
    def getUserID(self):
        return self.userID

class Book:
    def __init__(self, title):
        self.title = title
        self.borrowed = False
    def getTitle(self):
        return self.title
    def isBorrowed(self):
        return self.borrowed
    def setBorrowed(self, borrowed):
        self.borrowed = borrowed
\end{lstlisting}
\caption{A Python code obtained from the prompt in Figure \ref{fig:prompt1}}
\label{fig:python1}
\end{figure}

\subsection{Exploring the Use of ChatGPT for Bidirectional Traceability Problem}

\begin{figure}[h]
    \begin{tabularx}{\linewidth}{|X|}
    \hline
    \small
Isn't it true that ChatGPT cannot train models? Please suggest a method to synchronize without training the model.
	\\
    \hline
    \end{tabularx}
\caption{Prompt regarding the answer shown in Figure \ref{fig:answer1}}
\label{fig:requestion1}
\end{figure}

\begin{figure}[h]
\begin{tabularx}{\linewidth}{|X|}
\hline
\small
You're correct, ChatGPT itself cannot train models. It's a language model that generates human-like text based on the input it's given. 
\\
\\
\small
For bidirectional traceability without training a model, you can follow these steps:
\\
\\
\small
1. **Documentation**: Document the relationship between the design model and the code. This can be as simple as a comment in the code indicating which part of the design model it corresponds to, or a more complex document mapping each element of the design model to the corresponding code.
\\
\\
\small
2. **Consistent Naming Conventions**: Use consistent naming conventions in both your design model and code. This will make it easier to identify which parts of the code correspond to which parts of the design model.
\\
\\
\small
3. **Version Control**: Use a version control system like Git. This allows you to track changes in both your design model and code, making it easier to keep them in sync.
\\
\\
\small
4. **Regular Reviews**: Conduct regular reviews of both your design model and code to ensure they are still in sync. This could be part of a regular development cycle, such as a sprint review in Agile development.
\\
\\
\small
5. **Automated Testing**: Write tests to ensure that changes in the code still meet the requirements set out in the design model. This can help catch any discrepancies between the two.
\\
\\
\small
6. **Tools**: Use tools that support bidirectional traceability. Some integrated development environments (IDEs) and Unified Modeling Language (UML) tools support this feature.
\\
\\
\small
Remember, the key to maintaining bidirectional traceability is communication and consistency. Everyone on the team should understand the relationship between the design model and the code, and changes to either should be communicated clearly.
\\
\hline
\end{tabularx}
\caption{ChatGPT response to the prompt shown in Figure \ref{fig:requestion1}}
\label{fig:answer2}
\end{figure}


Responses to the prompt shown in Figure \ref{fig:prompt1} are presented in Figures \ref{fig:answer1}, \ref{fig:class1}, and \ref{fig:python1}. Although adequate results were obtained for design models and code, a satisfactory response for the bidirectional traceability problem using ChatGPT was not received. Therefore, we designed a new prompt (Figure \ref{fig:requestion1}) to query again. ChatGPT was asked about its inability to train the model and suggested a method to synchronize design models and code without training the model. The response received is shown in Figure \ref{fig:answer2}. This provides a known solution to the bidirectional traceability problem, but it does not consider the use of ChatGPT.

We asked multiple similar questions about using ChatGPT because we might get the intended result, but received no satisfactory responses.

\subsection{Testing Synchronization Check and Correction with ChatGPT}
Although it wasn't possible to determine the synchronization and correction method between design models and code by using ChatGPT, we thought it might be possible.

\begin{figure}[h]
    \begin{tabularx}{\linewidth}{|X|}
    \hline
    \small
Check if the changes between design models and Python code are synchronized, and if there are inconsistencies, propose corrections for both the design models and Python code.
\\
----
\\
\small
\#Design Model in PlantUML:
\\
\small
(the design model shown in Figure \ref{fig:class2})
\\
\\
\small
\#Python Code:
\\
\small
(the python code shown in Figure \ref{fig:python2})
	\\
    \hline
    \end{tabularx}
\caption{Prompt regarding synchronization check and propose corrections between design models and code}
\label{fig:check1}
\end{figure}

\begin{figure}[t]
\lstset{
    frame=single,
    tabsize=1,
    basicstyle=\ttfamily\footnotesize,
    frame=single,
    breaklines=true,
    numbers=left
}
\begin{lstlisting}[language=Python, numbers=left]
     :
```plantuml
@startuml
     :
class User {
  -namae: String
  -userCard: UserCard
  +User(namae: String)
  +getNamae(): String
  +getUserCard(): UserCard
}

class UserCard {
  -userID: String
  +UserCard(userID: String)
  +getUserID(): String
}
     :
@enduml
     :
\end{lstlisting}
\caption{A PlantUML class diagram as design model with ``name" changed to ``namae" and lines 6, 8, and 9 based on the code in Figure \ref{fig:class1}}
\label{fig:class2}
\end{figure}

\begin{figure}[t]
\lstset{
    frame=single,
    tabsize=1,
    basicstyle=\ttfamily\footnotesize,
    frame=single,
    breaklines=true,
    numbers=left
}
\begin{lstlisting}[language=Python]
     :
class User:
    def __init__(self, name, userCard):
        self.name = name
        self.userCard = userCard

    def getName(self):
        return self.name

    def getUserCard(self):
        return self.userCard

class UserCard:
    def __init__(self, userID:int):
        self.userID = userID

    def getUserID(self):
        return self.userID
     :
\end{lstlisting}
\caption{A Python code with ``:int" added on line 14 based on the code in Figure \ref{fig:python1}}
\label{fig:python2}
\end{figure}

\begin{figure}[h]
\lstset{
    frame=single,
    tabsize=1,
    basicstyle=\ttfamily\footnotesize,
    frame=single,
    breaklines=true
}
\begin{lstlisting}[]
The design model and Python code are mostly synchronized, but there are a few inconsistencies:

1. In the design model, the `User` class has a constructor (`User(namae: String)`) that takes one argument, but in the Python code, the `User` class has a constructor (`\_\_init\_\_(self, name, userCard)`) that takes two arguments.

2. In the design model, the `User` class has a `getNamae()` method, but in the Python code, the `User` class has a `getName()` method.

3. In the design model, the `UserCard` class has a constructor (`UserCard(userID: String)`) that takes a string argument, but in the Python code, the `UserCard` class has a constructor (`\_\_init\_\_(self, userID: int)`) that takes an integer argument.

Proposed corrections:

1. Update the design model's `User` class to include `userCard` in the constructor:

```plantuml
class User {
  -name: String
  -userCard: UserCard
  +User(name: String, userCard: UserCard)
  +getName(): String
  +getUserCard(): UserCard
}
```

2. Update the design model's `UserCard` class to reflect that `userID` is of type `int`:

```plantuml
class UserCard {
  -userID: int
  +UserCard(userID: int)
  +getUserID(): int
}
```

3. Update the Python code's `UserCard` class to reflect that `userID` is of type `String`:

```python
class UserCard:
    def __init__(self, userID:str):
        self.userID = userID

    def getUserID(self):
        return self.userID
```

Choose either correction 2 or 3 depending on whether you want `userID` to be a `String` or an `int`.
\end{lstlisting}
\caption{ChatGPT response to the prompt shown in Figure \ref{fig:check1}}
\label{fig:answer4}
\end{figure}

We designed a new prompt shown in Figures \ref{fig:check1}, \ref{fig:class2}, and \ref{fig:python2}, which was based on mutually consistent the design model shown in Figure \ref{fig:class1} and the Python code shown in Figure \ref{fig:python1}, each with some changes.
Lines 6, 8, and 9 of the PlantUML class diagram shown in Figure \ref{fig:class2} are the parts that changed from Figure \ref{fig:class2} to the intended change in the variable name ``name" to ``namae." 
Line 14 of the Python code shown in Figure \ref{fig:python2}, where ``:int" is added as the ``userID" variable type, is the part that was changed from Figure \ref{fig:python1}.

As a result, responses were received as shown in Figure \ref{fig:answer4}. These responses accurately reflected the intended changes, and the suggested corrections were also appropriate.

\section{Discussion}
\subsection{Generation of Design Models and Code from Natural Language Requirements Using ChatGPT}
The design model for the given requirements in Figure \ref{fig:prompt1} was created using PlantUML, accurately reflecting the requirements, as shown in Figure \ref{fig:class1}. Additionally, the Python code was generated, as shown in Figure \ref{fig:python1}. This process confirmed that ChatGPT can generate design models and code from natural language requirements, at least in the case of simple content. However, whether accurate design models and code can be generated for complex and numerous requirements remains unclear.

\subsection{Challenges and Limitations of Using ChatGPT for Bidirectional Traceability Problem Solving}
The proposed solution by ChatGPT to the bidirectional traceability problem demonstrates LLM's problem-solving capabilities. However, it was not successful in providing a solution using ChatGPT specifically. This is likely because ChatGPT hasn't learned the specific methods to address this problem.

Although Figures \ref{fig:answer1} and \ref{fig:answer2} in response to Figures \ref{fig:prompt1} and \ref{fig:requestion1} appear to show a method using ChatGPT, closer inspection reveals that the answer was inaccurate. This common phenomenon known as hallucination can occur when using an LLM (Language Model) \cite{hal}. It refers to the generation of non-existent information or data. GPT generates responses based on the data it has learned in the past. However, since ChatGPT has not known a specific solution to the bidirectional traceability problem in the past, the answer was likely created based on other ordinary methods, not ChatGPT.

\subsection{Assessing the Capability of ChatGPT in Implementing Bidirectional Traceability Between Design Models and Code}
Although ChatGPT could not provide a direct method using ChatGPT itself for implementing bidirectional traceability between specific design models and code, it could check the consistency between design models and the code. The ChatGPT could suggest corrections by using prompts from Figures \ref{fig:check1}, \ref{fig:class2}, and \ref{fig:python2}, as shown in Figure \ref{fig:answer4}. This demonstrated that ChatGPT can enable bidirectional traceability between design models and code.

However, this example was relatively simple, consisting of class diagrams and Python code. Therefore, further checking is required to assess how well ChatGPT can handle more complex inconsistencies on a larger scale. Additionally, there are different types of design models, such as sequence diagrams and communication diagrams. Checking whether ChatGPT can ensure consistency between these various models is also necessary.

\subsection{Use of Different Natural and Computer Languages in ChatGPT}
The use of Japanese and Java, instead of English and Python, was explored in a previous case study \cite{kanuka}. The results showed almost no difference in solving the problem with ChatGPT, regardless of whether natural languages or computer languages were used.

\section{Conclusion}
In the case study, the potential of ChatGPT using LLM GPT-4 was demonstrated to address software modeling challenges, particularly bidirectional traceability between design models and code. The study confirmed that ChatGPT can generate design models and code from natural language requirements. The results demonstrated ChatGPT's ability to effectively bridge the gap between natural language requirements and software modeling.

ChatGPT showed its problem-solving abilities by proposing ordinary solutions to bidirectional traceability. However, it was unable to offer solutions using specific methods by ChatGPT itself, indicating a limitation in its advanced problem-solving capabilities. This is likely because ChatGPT has not learned particular methods for solving the problem.

Despite this limitation, ChatGPT demonstrated consistency between design models and the code and suggested corrections, proving that bidirectional traceability between design models and code is achievable using ChatGPT.

\appendices
\section{Generating Design Model and Code}
We hypothesize that querying ChatGPT with prompts that subdivide the bidirectional traceability problem shown in Figure \ref{fig:prompt1} may yield more informative responses. We aim to test this hypothesis through the following extra case studies.

\subsection{Generating a Design Model by ChatGPT}
We designed a prompt (Figure \ref{fig:prompt10}) based on Figure \ref{fig:prompt1} to generate a design model. We had been asked to generate a design model in advance but obtained a sequence diagram instead of a class diagram. Therefore, `\#Instruction' in the prompt (Figure \ref{fig:prompt10}) instructed to generate a class diagram as a design model to ensure the correct one was generated.

Figure \ref{fig:answer10} generated by ChatGPT shows a class diagram in PlantUML that accurately represents the requirements mentioned in ``\#Problem." The class diagram (Figure \ref{fig:answer10}) has more extracted classes and their associations compared to the class diagram (Figure \ref{fig:class1}) while presenting almost the same amount of information. Figure \ref{fig:uml2} shows the class diagram that was generated from Figure \ref{fig:answer10}.

\begin{figure}[t]
\begin{tabularx}{\linewidth}{|X|}
\hline
\small
\#Problem:
\\
\small
(the text shown in Figure \ref{fig:prompt1})
\\
\\
\small
\#Instruction:
\\
\small
For the above \#problem, create \textbf{the class diagram} in PlantUML format in detail.
\\
\hline
\end{tabularx}
\caption{Prompt on generating a design model}
\label{fig:prompt10}
\end{figure}

\begin{figure}[t]
\lstset{
    frame=single,
    tabsize=1,
    basicstyle=\ttfamily\footnotesize,
    frame=single,
    breaklines=true
}
\begin{lstlisting}[]
@startuml
class Library {
  +openShelf()
  +closeShelf()
}

class User {
  +selectBook()
  +returnBook()
}

class Book {
  +getBookInfo()
}

class UserCard {
  +getUserInfo()
}

class CounterStaff {
  +registerLendingInfo()
  +performReturnProcess()
  +checkLendingStatus()
  +urgeDelayedUsers()
}

class LendingInformation {
  +getLendingInfo()
  +updateLendingInfo()
}

Library -- User : has >
Library -- CounterStaff : has >
User -- UserCard : has >
User -- Book : borrows >
UserCard -- CounterStaff : gives >
Book -- CounterStaff : gives >
CounterStaff -- LendingInformation : updates >
@enduml
\end{lstlisting}
\caption{ChatGPT response to the prompt shown in Figure \ref{fig:prompt10}}
\label{fig:answer10}
\end{figure}

\begin{figure}[h]
    \centering
    \includegraphics[width=6cm]{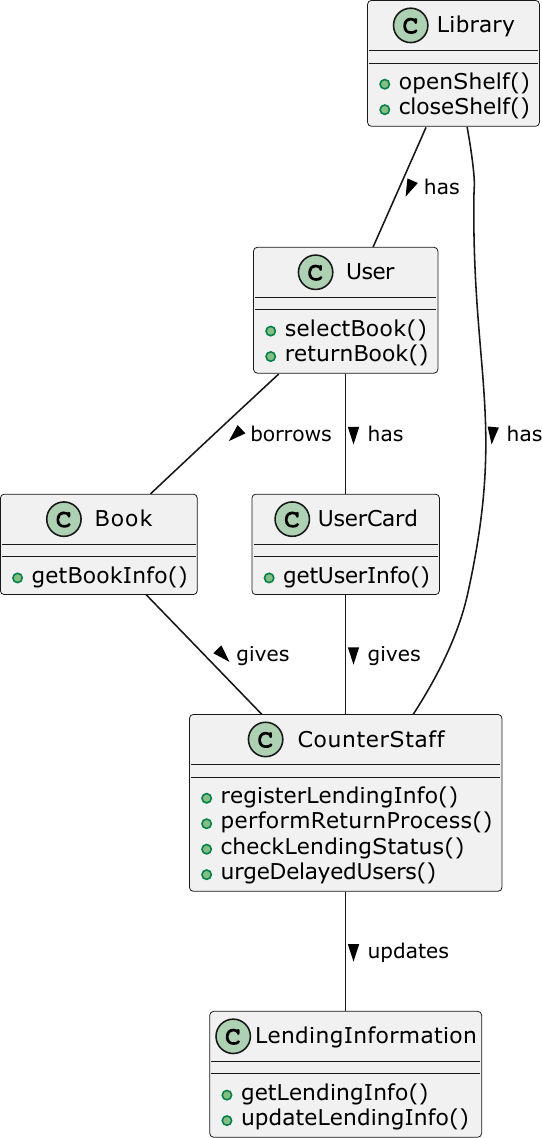}
    \caption{An image of the class diagram generated from Figure \ref{fig:answer10}}
    \label{fig:uml2}
\end{figure}

\subsection{Generating a Code by ChatGPT}

We designed a prompt (Figure \ref{fig:prompt11}) based on Figure \ref{fig:prompt1}, which was used to generate the Python code shown in Figure \ref{fig:answer11}. 

A comparison with the Python code (Figure \ref{fig:python1}) indicates that the Python code (Figure \ref{fig:answer11}) includes detailed business logic that accurately reflects the requirements, with specific methods such as ``lend\_book", ``return\_book", and ``check\_overdue\_books." The generated Python code also includes code examples that take use-case scenarios into consideration. Furthermore, additional comments for the generated code were also generated at the beginning and end of Figure \ref{fig:answer11}. This type of comment, not generated in Figure \ref{fig:python1}, is helpful information for users in handling and understanding the generated code.

\begin{figure}[h]
\begin{tabularx}{\linewidth}{|X|}
\hline
\small
\#Problem:
\\
\small
(the text shown in Figure \ref{fig:prompt1})
\\
\\
\small
\#Instruction:
\\
\small
For the above \#problem, create the code in python language in detail.
\\
\hline
\end{tabularx}
\caption{Prompt on generating a Python code}
\label{fig:prompt11}
\end{figure}

\begin{figure*}[h]
\lstset{
    frame=single,
    tabsize=1,
    basicstyle=\ttfamily\footnotesize,
    frame=single,
    breaklines=true
}
\begin{lstlisting}[]
Here is a simple implementation of the problem in Python. This code does not include any database or GUI interactions, it's just a simple console application.

```python
import datetime

class Book:
    def __init__(self, title, author):
        self.title = title
        self.author = author
        self.borrower = None
        self.borrow_date = None

class User:
    def __init__(self, name, phone):
        self.name = name
        self.phone = phone
        self.borrowed_books = []

class Library:
    def __init__(self):
        self.books = []
        self.users = []

    def add_book(self, title, author):
        self.books.append(Book(title, author))

    def add_user(self, name, phone):
        self.users.append(User(name, phone))

    def lend_book(self, user_name, book_title):
        for book in self.books:
            if book.title == book_title and book.borrower is None:
                for user in self.users:
                    if user.name == user_name:
                        book.borrower = user
                        book.borrow_date = datetime.datetime.now()
                        user.borrowed_books.append(book)
                        return "Book borrowed successfully"
        return "Book is not available"

    def return_book(self, user_name, book_title):
        for book in self.books:
            if book.title == book_title and book.borrower is not None:
                for user in self.users:
                    if user.name == user_name:
                        book.borrower = None
                        book.borrow_date = None
                        user.borrowed_books.remove(book)
                        return "Book returned successfully"
        return "Book is not available"

    def check_overdue_books(self):
        for user in self.users:
            for book in user.borrowed_books:
                if (datetime.datetime.now() - book.borrow_date).days > 14:
                    print(f"User {user.name} with phone number {user.phone} has an overdue book: {book.title}")

library = Library()
library.add_book("Book1", "Author1")
library.add_user("User1", "1234567890")
print(library.lend_book("User1", "Book1"))
print(library.return_book("User1", "Book1"))
library.check_overdue_books()
```

This code creates a `Book` class, a `User` class, and a `Library` class. The `Library` class has methods to add books and users, lend books, return books, and check for overdue books. The `lend_book` and `return_book` methods update the `borrower` and `borrow_date` attributes of the `Book` class and the `borrowed_books` attribute of the `User` class. The `check_overdue_books` method checks if any books have been borrowed for more than 14 days and prints a message if they have.
\end{lstlisting}
\caption{ChatGPT response to the prompt shown in Figure \ref{fig:prompt11}}
\label{fig:answer11}
\end{figure*}

\clearpage

\section{synchronizing design model and code by chatgpt}
ChatGPT could generate detailed design models (Figure \ref{fig:answer10}) and the Python code (Figure \ref{fig:answer11}). As these were generated individually, they can be inconsistent between them. Therefore, we designed a prompt (Figure \ref{fig:check10}) to check the consistency and propose appropriate correction methods for inconsistencies.

\begin{figure}[h]
    \begin{tabularx}{\linewidth}{|X|}
    \hline
    \small
Check if the changes between the design model and Python code are synchronized, and if there are inconsistencies, propose corrections for both design models and Python code.
\\
----
\\
\small
\#Design Model in PlantUML:
\\
\small
(the design model shown in Figure \ref{fig:answer10})
\\
\\
\small
\#Python Code:
\\
\small
(the python code shown in Figure \ref{fig:answer11})
	\\
    \hline
    \end{tabularx}
\caption{Prompt to check the consistency and propose corrections between design models and code}
\label{fig:check10}
\end{figure}

The ChatGPT response shown in Figure \ref{fig:answer14} presents the results that have the inconsistencies found in some items and their respective corrections proposed. Subsequently, we could have ChatGPT generate a revised design model (Figure \ref{fig:answer15}) and a revised Python code (Figure \ref{fig:answer16}) based on those correction approaches. Additionally, Figure \ref{fig:uml3} shows a class diagram image generated from the design model (Figure \ref{fig:answer15}).

The generated Python code in Figure \ref{fig:answer16} could be executed without any errors. However, ChatGPT might misunderstand the meaning of ``open-shelf" in the ``\#Problem" section and, as a result, generate an empty ``OpenShelf" method and ``CloseShelf" method in the Python code. The ``open-shelf" means picking up books directly from the library bookshelf and viewing them freely.

In conclusion, these extra case studies have taught us that it is better to generate design models and code individually and then check and correct their consistency. This approach allowed us to obtain more detailed design models and code. Furthermore, we discovered that ChatGPT can handle more complex consistency checks and corrections than we presented in this paper's main body.

\begin{figure}[h]
\begin{tabularx}{\linewidth}{|X|}
\hline
\small
The Python code does not fully reflect the design model. Here are the inconsistencies and proposed corrections:
\\
\\
\small
1. The design model includes a `UserCard` class, but the Python code does not. This class should be added to the Python code.
\\
\\
\small
2. The design model includes a `CounterStaff` class, but the Python code does not. This class should be added to the Python code.
\\
\\
\small
3. The design model includes a `LendingInformation` class, but the Python code does not. This class should be added to the Python code.
\\
\\
\small
4. The design model shows that the `User` class has a `selectBook()` and `returnBook()` methods, but the Python code does not. These methods should be added to the `User` class in the Python code.
\\
\\
\small
5. The design model shows that the `Book` class has a `getBookInfo()` method, but the Python code does not. This method should be added to the `Book` class in the Python code.
\\
\\
\small
6. The design model shows that the `Library` class has `openShelf()` and `closeShelf()` methods, but the Python code does not. These methods should be added to the `Library` class in the Python code.
\\
\\
\small
7. The design model shows that the `CounterStaff` class has `registerLendingInfo()`, `performReturnProcess()`, `checkLendingStatus()`, and `urgeDelayedUsers()` methods, but the Python code does not. These methods should be added to the `CounterStaff` class in the Python code.
\\
\\
\small
8. The design model shows that the `LendingInformation` class has `getLendingInfo()` and `updateLendingInfo()` methods, but the Python code does not. These methods should be added to the `LendingInformation` class in the Python code.
\\
\\
\small
9. The design model shows that the `UserCard` class has a `getUserInfo()` method, but the Python code does not. This method should be added to the `UserCard` class in the Python code.
\\
\\
\small
10. The design model shows relationships between classes that are not reflected in the Python code. These relationships should be added to the Python code.
\\
\\
\small
In the design model, the `User` class should not directly interact with the `Book` class. Instead, the `User` should interact with the `Library` or `CounterStaff` to borrow or return books. The `CounterStaff` should then update the `LendingInformation`. This change should be reflected in both the design model and the Python code.
\\
\hline
\end{tabularx}
\caption{ChatGPT response to the prompt shown in Figure \ref{fig:check10}}
\label{fig:answer14}
\end{figure}

\begin{figure}[h]
\lstset{
    frame=single,
    tabsize=1,
    basicstyle=\ttfamily\footnotesize,
    frame=single,
    breaklines=true
}
\begin{lstlisting}[]
Here is the updated design model in PlantUML:

```plantuml
@startuml
class Library {
  +openShelf()
  +closeShelf()
  +addBook()
  +addUser()
  +lendBook()
  +returnBook()
  +checkOverdueBooks()
}

class User {
  +selectBook()
  +returnBook()
}

class Book {
  +getBookInfo()
}

class UserCard {
  +getUserInfo()
}

class CounterStaff {
  +registerLendingInfo()
  +performReturnProcess()
  +checkLendingStatus()
  +urgeDelayedUsers()
}

class LendingInformation {
  +getLendingInfo()
  +updateLendingInfo()
}

Library -- User : has >
Library -- CounterStaff : has >
User -- UserCard : has >
User -- Library : interacts >
UserCard -- CounterStaff : gives >
Book -- CounterStaff : gives >
CounterStaff -- LendingInformation : updates >
@enduml
```

This updated design model now includes the methods that were missing in the original design model and reflects the changes proposed in the Python code. The relationship between the `User` and `Book` classes has been removed and replaced with a relationship between the `User` and `Library` classes. The `User` now interacts with the `Library` to borrow or return books. The `CounterStaff` then updates the `LendingInformation`.
\end{lstlisting}
\caption{ChatGPT response to the prompt shown in Figure \ref{fig:check10}}
\label{fig:answer15}
\end{figure}

\begin{figure}
    \centering
    \includegraphics[width=6cm]{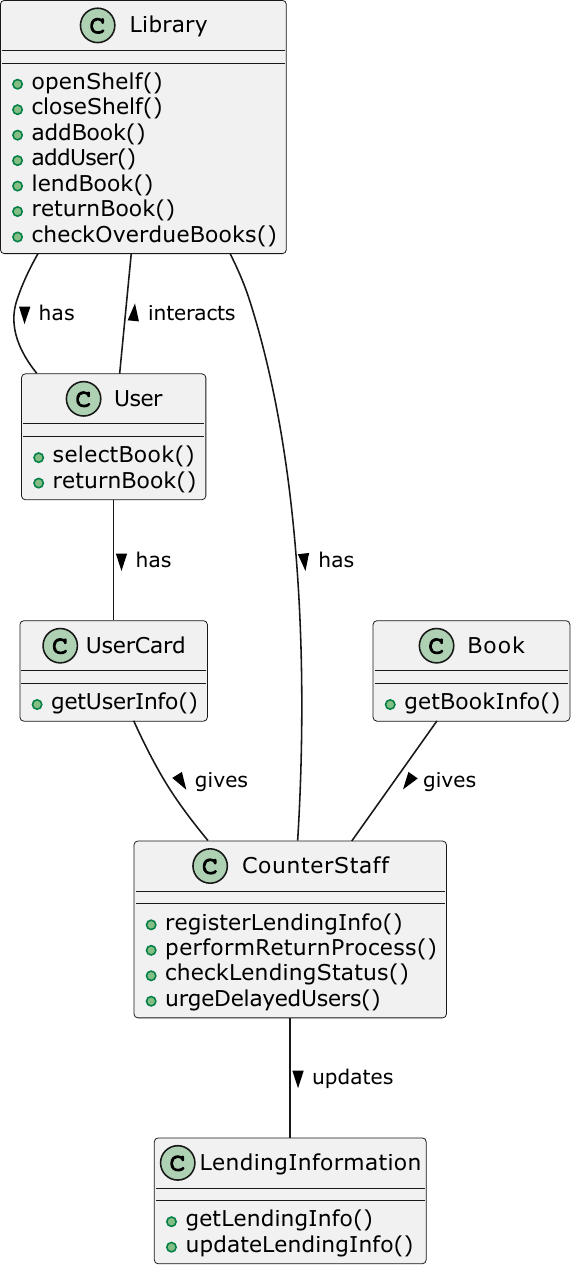}
    \caption{An image of the class diagram generated from Figure \ref{fig:answer15}}
    \label{fig:uml3}
\end{figure}

\begin{figure*}[h]
\lstset{
    frame=single,
    tabsize=1,
    basicstyle=\ttfamily\tiny,
    frame=single,
    breaklines=true
}
\begin{lstlisting}[]
Here is the updated Python code that reflects the changes proposed in the design model:

```python
import datetime

class Book:
    def __init__(self, title, author):
        self.title = title
        self.author = author
        self.borrower = None
        self.borrow_date = None
    def getBookInfo(self):
        return self.title, self.author

class User:
    def __init__(self, name, phone):
        self.name = name
        self.phone = phone
        self.borrowed_books = []
    def selectBook(self, library, book_title):
        library.lend_book(self.name, book_title)
    def returnBook(self, library, book_title):
        library.return_book(self.name, book_title)

class UserCard:
    def __init__(self, user):
        self.user = user
    def getUserInfo(self):
        return self.user.name, self.user.phone

class CounterStaff:
    def __init__(self):
        self.lending_info = {}
    def registerLendingInfo(self, user, book):
        self.lending_info[user.name] = book.title
    def performReturnProcess(self, user, book):
        del self.lending_info[user.name]
    def checkLendingStatus(self, user):
        return self.lending_info.get(user.name)
    def urgeDelayedUsers(self, library):
        library.check_overdue_books()

class LendingInformation:
    def __init__(self):
        self.info = {}
    def getLendingInfo(self, user):
        return self.info.get(user.name)
    def updateLendingInfo(self, user, book):
        self.info[user.name] = book.title

class Library:
    def __init__(self):
        self.books = []
        self.users = []
        self.staff = CounterStaff()
    def openShelf(self):
        pass
    def closeShelf(self):
        pass
    def add_book(self, title, author):
        self.books.append(Book(title, author))
    def add_user(self, name, phone):
        self.users.append(User(name, phone))
    def lend_book(self, user_name, book_title):
        for book in self.books:
            if book.title == book_title and book.borrower is None:
                for user in self.users:
                    if user.name == user_name:
                        book.borrower = user
                        book.borrow_date = datetime.datetime.now()
                        user.borrowed_books.append(book)
                        self.staff.registerLendingInfo(user, book)
                        return "Book borrowed successfully"
        return "Book is not available"
    def return_book(self, user_name, book_title):
        for book in self.books:
            if book.title == book_title and book.borrower is not None:
                for user in self.users:
                    if user.name == user_name:
                        book.borrower = None
                        book.borrow_date = None
                        user.borrowed_books.remove(book)
                        self.staff.performReturnProcess(user, book)
                        return "Book returned successfully"
        return "Book is not available"
    def check_overdue_books(self):
        for user in self.users:
            for book in user.borrowed_books:
                if (datetime.datetime.now() - book.borrow_date).days > 14:
                    print(f"User {user.name} with phone number {user.phone} has an overdue book: {book.title}")

library = Library()
library.add_book("Book1", "Author1")
library.add_user("User1", "1234567890")
user1 = library.users[0]
user1.selectBook(library, "Book1")
user1.returnBook(library, "Book1")
library.staff.urgeDelayedUsers(library)
```

This updated Python code now includes the `UserCard`, `CounterStaff`, and `LendingInformation` classes that were missing in the original Python code. The `User` class now interacts with the `Library` to borrow or return books. The `CounterStaff` then updates the `LendingInformation`.
\end{lstlisting}
\caption{ChatGPT response to the prompt shown in Figure \ref{fig:check10}}
\label{fig:answer16}
\end{figure*}


\begin{thebibliography}{10}

\bibitem{b1} OpenAI, ``ChatGPT," [Online]. Available: https://openai.com/chatgpt.

\bibitem{llm} X. Hou, Y. Zhao, Y. Liu, Z. Yang, K. Wang, L. Li, X. Luo, D. Lo, J. Grundy, and H. Wang, ``Large Language Models for Software Engineering: A Systematic Literature Review," arXiv preprint arXiv:2308.10620, 2023.

\bibitem{barke}
S. Barke, M. B. James, and N. Polikarpova, ``Grounded Copilot: How Programmers Interact with Code-Generating Models," Proc. ACM Programming Languages, vol. 7, no. OOPSLA1, pp. 85–111, 2023.

\bibitem{sakura}
T. Sakura, R. Soga, H. Kanuka, K. Shimari, T. Ishio, "Leveraging Execution Trace with ChatGPT: A Case Study on Automated Fault Diagnosis," in Proceedings of the 39th IEEE International Conference on Software Maintenance and Evolution (ICSME), pp. 397-402, 2023.

\bibitem{tasks} Z. Zheng, K. Ning, J. Chen, Y. Wang, W. Chen, L. Guo, and W. Wang, ``Towards an Understanding of Large Language Models in Software Engineering Tasks," arXiv preprint arXiv:2308.11396, 2023.

\bibitem{camara}
J. Cámara, J. Troya, L. Burgueño, and A. Vallecillo, ``On the assessment of generative AI in modeling tasks: an experience report with ChatGPT and UML," Software and Systems Modeling, pp. 1–13, 2023.

\bibitem{gpt4}
OpenAI, ``GPT-4 Technical Report," arXiv:2303.08774 [cs.CL], 2023.

\bibitem{trace}
F. Tian, T. Wang, P. Liang, C. Wang, A. A. Khan, and M. A. Babar, ``The impact of traceability on software maintenance and evolution: A mapping study," Journal of Software: Evolution and Process, vol. 33, no. 10, pp. e2374, 2021.

\bibitem{eramo}
R. Eramo, A. Pierantonio, and M. Tucci, ``Improved traceability for bidirectional model transformations," Proc. MODELS 2018 Workshops co-located with ACM/IEEE 21st International Conference on Model Driven Engineering Languages and Systems (MODELS 2018), vol. 2245, pp. 306-315.

\bibitem{ubayashi}
N. Ubayashi and Y. Kamei, ``Library problem 2.0: A common problem example in software engineering research," Proc. SIG Software Engineering (SIGSE) Winter Workshop 2012 in Biwako, Information Processing Society of Japan, pp. 129-130, 2012. (in Japanese)

\bibitem{plantuml}
``PlantUML," [Online]. Available: https://plantuml.com.

\bibitem{fewshot}
Prompting Guide, ``Few-shot Prompting Techniques," [Online]. Available: https://www.promptingguide.ai/techniques/fewshot.

\bibitem{hal}
Z. Ji, N. Lee, R. Frieske, T. Yu, D. Su, Y. Xu, E. Ishii, Y. J. Bang, A. Madotto, and P. Fung, ``Survey of Hallucination in Natural Language Generation," ACM Computing Surveys, vol. 55, no. 12, pp. 248-285, Mar. 2023.

\bibitem{kanuka}
H. Kanuka, G. Koreki, R. Soga, and K. Nishikawa, ``An experiment for applying ChatGPT to bidirectional traceability
problem between design models and code," Proc. SIG Software Engineering (SIGSE) Software Engineering Symposium 2023, Information Processing Society of Japan, pp. 238-239, Aug. 2023. (in Japanese)

\end{thebibliography}
\end{document}